# Spin-transfer-torque resonant switching and injection locking in presence of a weak external microwave field for spin valves with perpendicular materials

Mario Carpentieri[1], Giovanni Finocchio[2], Bruno Azzerboni[2], Luis Torres[3]

[1]University of Calabria, Via P. Bucci 42C, I-87036, Rende (CS), Italy.

[2]Department of Fisica della Materia e Ingegneria Elettronica. University of Messina, C.da di Dio, I-98166, Messina, Italy.

[3]Departamento de Fisica Aplicada, University of Salamanca, Plaza de la Merced s/n, E-37008 Salamanca, Spain.

**Abstract**: The effects of a weak microwave field in the magnetization dynamics driven by spin-transfer-torque in spin-valves with perpendicular materials have been systematically studied by means of full micromagnetic simulations. In the system we studied, depending on the working point (bias field and current) in the dynamical stability diagram, we observe either resonant switching and injection locking.

The resonant switching, observed in the switching region, occurs when the field frequency is approaching the frequency of the main pre-switching mode giving rise to an asymmetric power distribution of that mode in the sectional area of the free layer. At the resonant frequency, the switching time is weakly dependent on the relative phase between the instant when the current pulse is applied and the microwave field.

The injection locking, observed in the dynamical region, is characterized by the following properties: (i) a locking bandwidth which is linearly dependent on the force locking, and (ii) a locking for integer harmonics of the self-oscillation frequency. We compare our numerical results with analytical theory for non-autonomous non-linear system obtaining a good agreement in the current region where the oscillation frequency and output power are characterized from a linear relationship.

# I. INTRODUCTION

The control of the magnetization dynamics in ferromagnets at nanoscale is nowadays a key aspect for technological point of view. In particular, the possibility to manipulate the magnetization state by means of a spin-polarized current via spin-transfer torque (STT) opens perspectives for magnetic storage elements, nanoscale microwave sources, radio-frequency detectors and modulators.[1,2,3,4,5,6] The basic phenomena are for current flowing through in two nanomagnets separated by a normal metal spacer or a thin insulator, where one of the nanomagnet acts as "polarizer layer" and the other as "free layer".[7,8]

For data storage applications, STT-devices have to satisfy two main features: (i) reduced critical current in order to be compatible with the CMOS technology, and (ii) high thermal stability. Reductions of the critical current have been obtained with different strategies involving complicated structures.[9,10,11,12] A very interesting result is the possibility to have current induced magnetization reversal in spin valves with free and polarizer layer realized with perpendicular materials[13] where, it is possible to achieve the manipulation of a magnetic state with low current density (e.g. $7x10^6$ A/cm$^2$ in Ref. 14) maintaining high thermal stability. To further improve the switching properties in those spin-valves a resonant switching-based memory architectures can be a key solution.

For microwave source applications, the main limitations of spin-torque oscillators are their low output power and large linewidth. From experimental point of view to achieve enhanced emitted power and reduced linewidth, the simplest solution is the phase locking of array of spin-torque oscillators (STO).[15,16,17,18] It is very important from fundamental point of view, to study the way to control an array of STOs to a fixed controlled frequency by means of the injection locking mechanism (via a "weak" microwave source) in order to design practical STOs.[19,20] Experimentally, the injection locking due to a microwave current (locking bandwidth of tens of MHz) has been observed for point contact geometries[21] and spin valves.[22] It should be noted that, even if the injection locking with a microwave current is the simplest for the experimental realization, the locking bandwidth, being proportional to the tangent of the angle between the equilibrium orientations of the magnetization in the free and the polarizer layer, is sensibly reduced since this angle is usually small.[19] From a theoretical point of view a different driving signal as a microwave field it is preferred.[19,23]

Here, we performed a complete study by means of full micromagnetic simulations of the magnetization reversal and persistent oscillation of the magnetization driven by the simultaneous application of a weak linearly polarized ac field and dc spin-polarized current in spin valves with free and polarizer layer of perpendicular material.

In the first part of the paper, we will discuss the resonant switching. We find that as the field frequency approaches the resonant frequency the switching time is sensibly reduced, and at that frequency exists a transition from uniform to non-uniform spatial distribution of the main pre-switching excited mode. In other words, the ac field helps the switching because it speeds up the break of the symmetry of the pre-switching oscillations. Differently from the resonant switching driven by a microwave current, here we find out that at the resonant frequency the switching time is weakly dependent on the relative phase between the instant when the current pulse is applied and the ac field.

In the second part of the paper, we point our attention to the study of the injection locking mechanism.[24] We compare our numerical results with a non-linear analytical theory (no fitting parameters), where the oscillation power and the dimensionless non-linear frequency shift have been computed from the free running data.[19] We obtained a quantitative agreement for the locking bandwidth as function of the microwave field amplitude for current density near the threshold. We also found the injection locking phenomenon to the 2nd and 3rd harmonic. This harmonic-injection-locked oscillator can be used as an injection-locked frequency divider.

This paper is organized as follow, Section II describes the numerical details of the micromagnetic model used for our numerical experiment. In Sections III-V are discussed the effects of a microwave field on the switching processes (resonant switching) and on the persistent magnetization oscillation (injection locking).

## II. NUMERICAL DETAILS

We studied magnetization dynamics in spin valves with perpendicular material CoPt-CoNi(4.5nm) (polarizer)/Cu(4nm)/CoNi(2.8nm) (free layer) and elliptical cross sectional area (100nm x 50nm). Figure 1 shows a schematic description of the spin-valves under investigation. We introduce a Cartesian coordinate system with the $x$ and the $y$-axis related to the easy and the hard in-plane axis of the ellipse respectively. Our results are based on the numerical solution[25,26] of the Landau-Lifshitz-Gilbert-Slonczewski (LLGS) equation,[7] we included together with the standard effective field (external, exchange, self-magnetostatic, and perpendicular uni-axial anisotropy) the magnetostatic field due to the polarizer layer and the Oersted field due to the current density (for a complete numerical description of the model see [25] and [27]). The simulation parameters are: $500 \times 10^3$ A/m and $650 \times 10^3$ A/m saturation magnetization of polarizer and free layer, $20 \times 10^5$ J/m$^3$ and $3.3 \times 10^5$ J/m$^3$ perpendicular anisotropy constant of polarizer and free layer, exchange constant $1.2 \times 10^{-11}$ J/m, damping parameter 0.1, we use for the torque the expression computed by Slonczweski in 1996[7] with polarization of $P = 0.35$. Those parameters are the same of the ones used in the Ref.[13] to numerically reproduce the data of that experimental framework. We performed some simulations with damping 0.025 and $P$=0.4 obtaining qualitatively the same results described in the next sections.

For the resonant switching simulations, we apply a bias field with an amplitude of -46 mT in order to compensate the magnetostatic coupling with the polarizer (the magnetization of the polarizer is fixed along the +$z$-direction (0,0,1)). The external field was applied with a tilted angle of 5° with respect to the $z$ direction along the +$x$-axis direction, in order to control the in-plane component of the magnetization (see Fig. 1 $H_{app}$-switching). For the injection locking simulations, the bias field is applied along the $x$-direction (see Fig. 1 $H_{app}$-dynamics). For each field value, the magnetic configuration of the polarizer has been computed by solving the static LLG equation (magnetization parallel to the effective field) for the whole CoPt-CoNi/Cu/CoNi system.

The free layer has been discretized in computational cells of 2.5 x 2.5 x 2.8 nm$^3$. The time step used was 32 fs. The current density is considered positive when the electrons flow from the polarizer to the free layer. The thermal fluctuations have been taken into account as an additive stochastic contribution to the deterministic effective field for each computational cell.[28,29,30,31] The microwave field is linearly polarized at 45° of amplitude $h$ in the $x$-$y$ plane (see Eq. 1) and it is directly added to the effective field.

$$h_{AC} = h_x \sin(2\pi f_{AC} t + \phi_H)\hat{x} + h_y \sin(2\pi f_{AC} t + \phi_H)\hat{y} \tag{1}$$

The $t, f_{AC}$ and $\phi_H$ are the time, the field frequency and phase, $h_x = h_y = 0.5\sqrt{2}\ h$, $\phi_H = 45°$.[32]

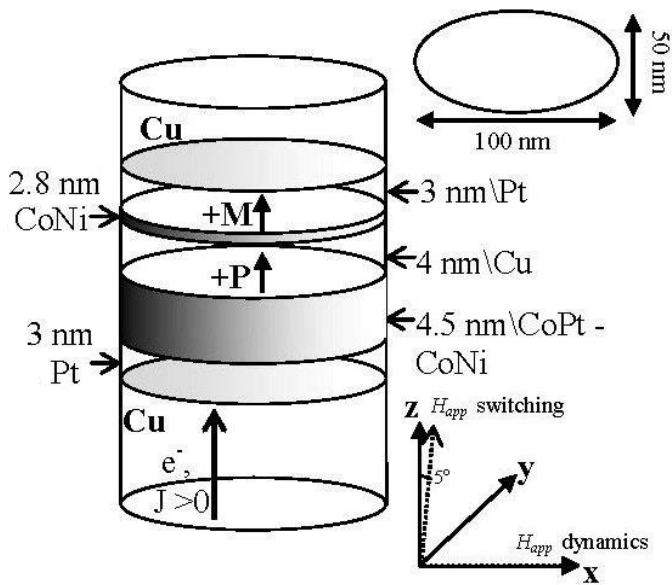

FIG. 1: Sketch of the simulated spin-valves. For the switching processes the external field is applied mainly along the $z$-direction $H_{app}$-switching, for the precessional dynamics the external field is applied along the $x$-direction $H_{app}$-dynamics.

## III. RESONANT SWITCHING PROCESSES

Here, we systematically studied results of resonant switching for fast switching processes with reversal time smaller than 3.5 ns (STT switching regime).[33] First of all, we characterized the switching processes with no microwave field, figure 2(a) displays the reversal time[34] computed for the P→AP (parallel to anti-parallel configuration of magnetization of the polarizer and the free layer) and AP→P transitions for $T$=0 K and $T$=300 K as function of the current density. For $T$=300 K we consider the reversal time as the most probable in the distribution of the reversal time computed for 100 realizations (see the inset of Fig. 2(a)), we also included some error bars to have a better idea of the distribution width (computed at the values where the switching probability is 50% of the maximum value) as function of current density. As observed in Fig. 2(a), to have the reversal time the current density for the P→AP transition has to be at least two times larger than the one for the AP→P transition (due to asymmetry of the STT). An example of a reversal time trace for P→AP transition is displayed in Fig. 2(b) for a current density of $J$= -8.5x10$^7$ A/cm$^2$ ($T$=0 K).

Our data show different qualitative behaviour for the reversal processes. For the AP→P the switching time is almost independent of the thermal fluctuations, while for the P→AP the switching time depends on the thermal fluctuations (compare the data in Fig. 2(a)).[35] The origin of this difference is due to the Oersted field, in fact it plays a more important role in P→AP transitions due to the larger current necessary to obtain the switching. This conclusion is also supported by simulations performed with no Oersted field (not shown) where also for the P→AP transitions the switching time is independent of the thermal fluctuations.[36] Qualitatively similar results have been also observed in the experiments of Ref.[35].

The micromagnetic spectral analysis[37,38] of the small oscillation region from the time $t_0$ (when the current pulse reaches its maximum value (rise time 0.05 ns)) up to $t_1$ ($t_0$ + 1 ns) shows that two pre-switching modes are excited during the fast switching processes (see Fig. 2(c)). For the P→AP transition, the lower frequency mode (P$_1$, $f$=8.8 GHz)[39] has much larger power than the higher frequency mode (P$_2$, $f$=10.3 GHz) (Fig. 2(c) bottom curve), differently for the AP→P transition the power of the modes AP$_1$ ($f$=9.0 GHz) and AP$_2$ ($f$=11.3 GHz) is substantially the same (Fig. 2(c) top curve). The excited modes are edge modes distributed symmetrically in the cross sectional area of the spin-valve, the spins oscillate out-of phase giving rise to a giant-magneto-resistance (GMR) signal with very small oscillations (see Fig. 2(b)). Another interesting feature of our data, it is that the frequencies of those excited modes are almost independent of the amplitude of the current density pulse in the range -11x10[7] A/cm$^2$<$J$<-8x10[7] A/cm$^2$ for P➔AP transition and 2x10[7] A/cm$^2$<$J$<4x10[7] A/cm$^2$ for AP➔P transition, even if the thermal fluctuations (300 K) are taken into account.

Experimentally, the effect of a microwave current on the critical switching current[40] and resonant switching[41] has been already studied. Particularly in Ref. [41], it is measured that STT from a microwave-frequency current pulse can produce a resonant excitation of a nanomagnet and improve fast switching characteristics in combination with a square current pulse. The larger speed-up of the switching time occurs at the resonant frequency if the phase of the microwave current at the onset time of the square pulse is the optimum value. In that experiment a strong dependence of the switching time on that phase value has been found.

The inset of Fig. 2(b) shows an example of our excitation waveform which combines a microwave field and a square-wave current density pulse, the phase $\varphi_D$, between the microwave field and the instant when the current pulse is applied, is computed from the time delay $t_D$ as $\varphi_D = 2\pi f_{AC} t_D$. In the rest of the paper, we show results due to a microwave field amplitude of $h$=1 mT, similar results have been obtained for $h$=0.5 and 2 mT. Figures 2(d) and (e) show the switching time as function of the field frequency ($T$=0 K,) for two different current densities $J$=-8.5x10[7] A/cm$^2$ and $J$=2.7x10[7] A/cm$^2$ for the P➔AP and AP➔P magnetization reversal respectively. For the P➔AP transition, the switching time reaches a minimum at the field frequency of the P$_1$ mode and for field frequency larger than 11.0 GHz the switching time stabilizes to a value larger than the one at zero ac field. Basically, around that frequency, the spatial distribution of the P$_1$ mode computed by means of the micromagnetic spectral mapping technique becomes asymmetric (see insets of Fig. 2(d)) favouring the switching process. The spatial distribution of the P$_2$ mode does not change for any frequency of the microwave field. A similar result is observed for the AP➔P transition, in particular a minimum switching time is achieved at the field frequency of the AP$_1$ mode, at that frequency also the spatial distribution of the AP$_1$ becomes asymmetric (see insets of Fig. 2(e)). For the AP➔P switching processes, the dependence of the switching time as function of the field frequency is strongly non linear, and its value tends to be constant for a large field frequency (as example for the data in Fig. 2(e) this value is 14 GHz). We also found that the spatial distribution of AP$_2$ mode does not change for any field frequency. Those trends are qualitatively similar for other current densities, -11x10[7] A/cm$^2$<$J$<-8x10[7] A/cm$^2$ for P➔AP transition and 2x10[7] A/cm$^2$<$J$<4x10[7] A/cm$^2$ for AP➔P transition. The microwave field does not affect the switching mechanism for the other current density range.

Differently from the resonant switching assisted by a microwave current, our numerical results show at the resonant frequency a switching time almost independent of the phase $\varphi_D$. In particular, for $J$=-8.5x10$^7$ A/cm$^2$, by changing $\varphi_D$ by step of 15°, we find that the main difference between the maximum and the minimum switching time was less than 0.2 ns ($f_{AC}$= 8.8 GHz). The origin of this behaviour is related to the fact that the conservative torque due to the microwave field drives the magnetization of the free layer to oscillate in a circular trajectory around its equilibrium axis, as consequence for systems with collinear configuration of the equilibrium axis of the free layer and the polarizer, the initial STT due to the current density pulse will be independent on the time of application (the angle between the two magnetizations is constant in time). When the free layer equilibrium axis and the magnetization of the polarizer are not collinear (as in our system due to the misalignment introduced by the external dc field), this gives rise to a weakly switching time dependence on the relative phase $\varphi_D$. Simulations performed with an external field applied with a misalignment of 1° show the switching time dependence on $\varphi_D$ completely disappears.

Figure 2(f) summarizes results of the switching time as function of the current density for $T$=0 K and $T$=300 K at the resonant frequency. As can be observed, the gap in the switching time for the P→AP due to the thermal fluctuations is deleted, and a narrow distribution of the switching time as function of the number of switching processes is achieved (see inset of Fig. 2(f)).

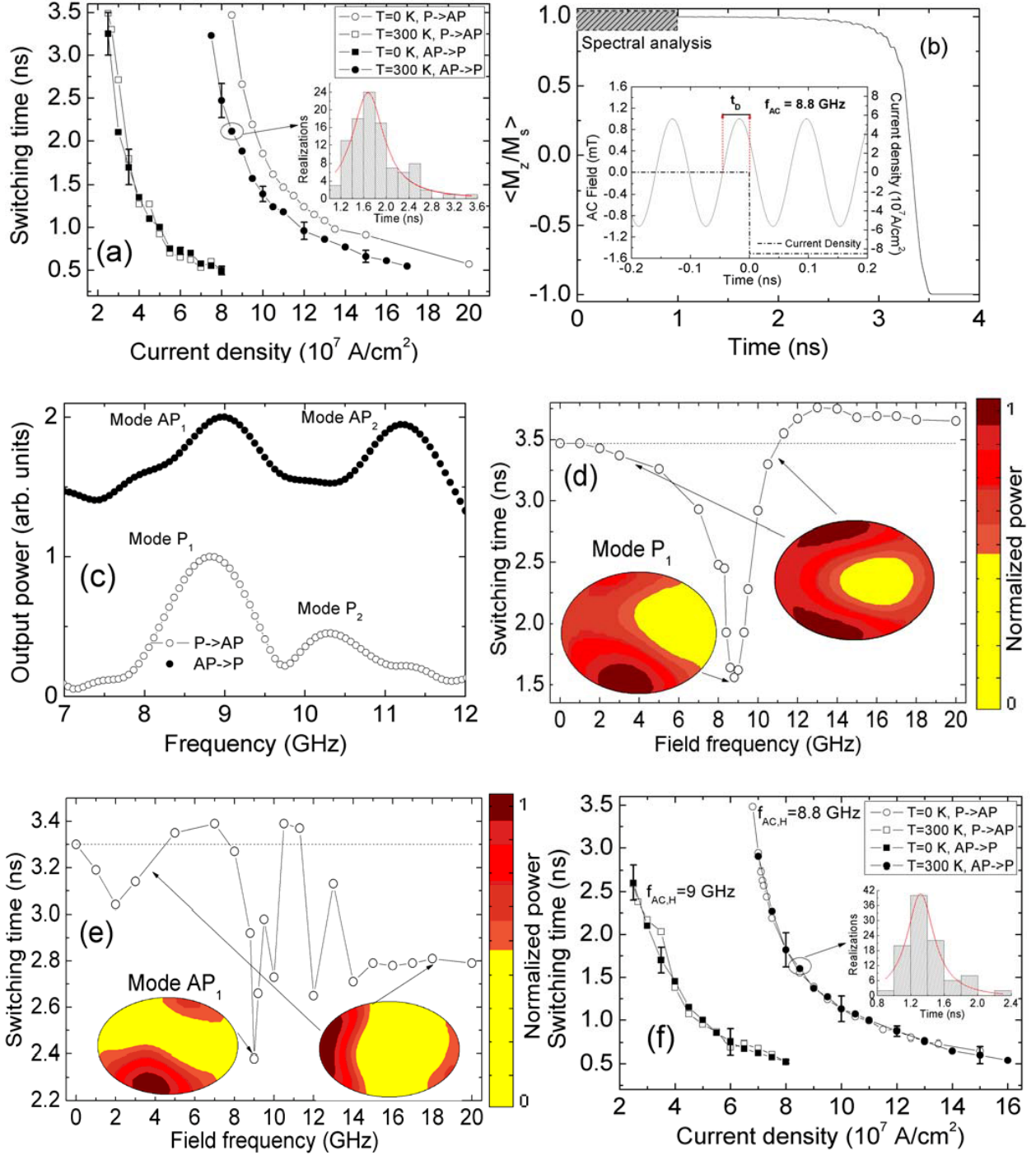

FIG. 2. (Color online) (a) Switching time for P→AP and AP→P transitions for $T$=0 K and $T$=300 K as function of the current density. Inset: probability histogram for 100 realizations at $T$=300 K. (b) Temporal evolution of the magnetization for P→AP transition at a current density of $J$= -8.5x10$^7$ A/cm$^2$. Inset: example of excitation waveform which combines microwave field and square-wave current density pulse. (c) Frequency spectra for the first nanosecond of reversal for P→AP (bottom curve, $J$=-8.5x10$^7$ A/cm$^2$) and AP→P (top curve, $J$=2.7x10$^7$ A/cm$^2$) transitions (an offset is applied). (d) and (e) show the switching time as function of the field frequency together with the spatial distribution of the main excited mode for P➔AP and AP➔P transitions respectively. (f) Switching time as function of the current density at *T*=0 K and *T*=300 K for resonant switching assisted by a microwave field. Inset: probability histogram for 100 realizations at *T*=300 K.

## IV. PERSISTENT OSCILLATIONS

Before to study the injection locking mechanism, we identify the dynamical region of dynamical stability diagram. In those spin-valves, we do not observe precessional dynamics for out-of-plane fields (up to 750 mT along *z*-direction) basically because the energy landscape is mainly characterized by two minima related to the parallel and anti-parallel configuration. For large in-plane fields, at *T*=0 K, we found self-oscillations (free running data) with a single-excited mode (linewidth limited only by the simulation time) in a wide range of current density. In Fig. 3, it is displayed the precessional dynamical region (we also plot the threshold current densities as function of the in-plane bias field). The micromagnetic computations in that region will be used to identify the oscillation power (*T*=0 K) and the dimensionless non-linear frequency shift (*T*=300 K) which are necessary to compare our numerical results of injection locking with the prediction of a non-linear analytical theory.[19]

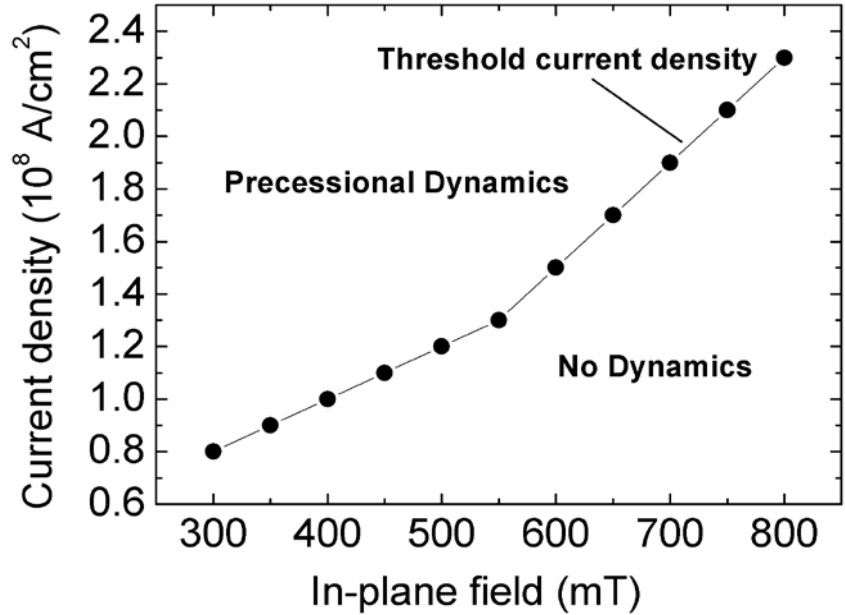

FIG. 3. Threshold current density as function of the in-plane bias field, and indication of the precessional dynamical region.

Even though here we describe in detail computations related to the bias field region between 450 and 550 mT, the results obtained for a wide range of field in the dynamical region are qualitatively similar. Figures 4(a) and (b) show the oscillation frequency $f$ and the dimensionless power $p$ computed as the non-linear power[42,43] as function of the current density for three external fields (450, 500, and 550 mT). As can be observed both oscillation power and frequency are non-monotonic function of the current density.[42] The parametric plot (the current density increases) in Fig. 4(c) shows two regions of linear relationship of the oscillation frequency as function of the oscillation power, which are characterized from two different values of non-linear frequency shift $N = 2\pi \dfrac{df}{dp}$,[19] for $H$=550 mT ($N_1$=18 GHz$^{-1}$, $N_2$=49.8 GHz$^{-1}$), for $H$=500 mT ($N_1$=18.1 GHz$^{-1}$, $N_2$=38.8 GHz$^{-1}$), and for $H$=450 mT ($N_1$=18.3 GHz$^{-1}$, $N_2$=37.6 GHz$^{-1}$).[19]

In the following part of the paper, we point out our attention to computations related to the external field $H$=500 mT, qualitatively similar results are also observed for other external fields. Figure 4(d) shows the linewidth $\Delta f$ for a temperature of $T$=300 K (computed by means of a Lorentzian fit of the power spectrum) as function of the current density (simulation time of 550 ns which fixes the minimum numerical linewidth to 1.8 MHz) compared to the analytical formula (the dimensionless nonlinear frequency shift $\nu$ is used as a fitting parameter) of Ref. [44]:

$$\Delta f = \Gamma_0(p)\frac{k_B T}{\beta p}\left(1+\nu^2\right) \qquad (2)$$

$\Gamma_0(p)$ is the natural positive damping computed for the dimensionless oscillation power $p$ (see Fig. 4(b)), $\beta = 2\pi f \dfrac{M_S V_{EFF}}{\gamma}$ is the coefficient relating the averaged energy to the dimensionless oscillation power, $V_{EFF}$ is the effective magnetic volume where the magnetization oscillates (in this system is the whole free layer volume), $k_B$ and $\gamma$ are the Boltzmann constant and the electron gyromagnetic ratio. We found for the term $\left(1+\nu^2\right)$ a value of 2.5. Experimentally, the term $\left(1+\nu^2\right)$ can also be estimated using the procedure presented in Ref. [45].

As can be observed for current density near the threshold ($J$<1.6x10$^8$A/cm$^2$), the analytical linewidth computed by Eq. 2 differs largely from the micromagnetic linewidth as expected for STOs with large non-linear frequency shift coefficient.[46]

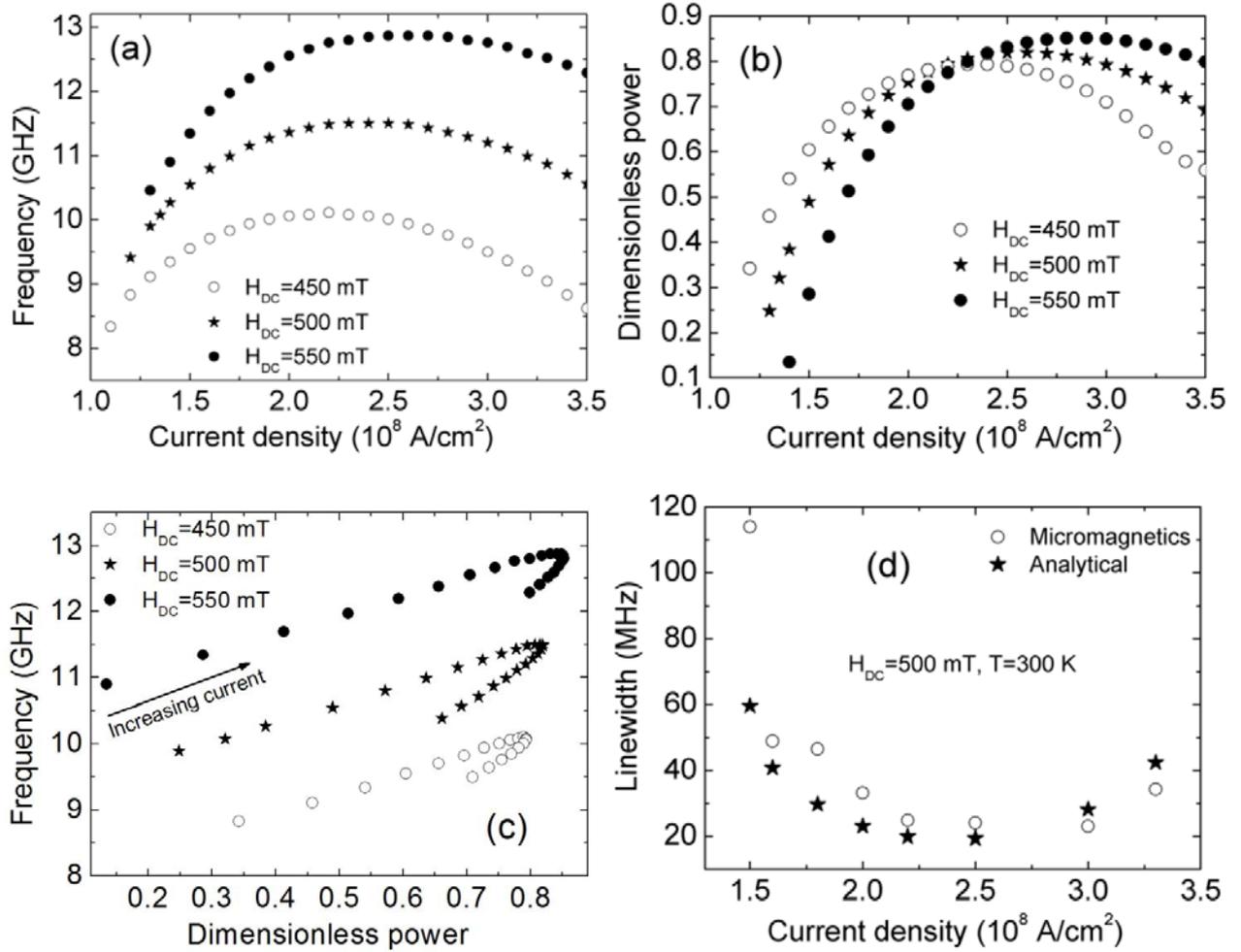

FIG. 4. (a) Oscillation frequency $f$ and (b) dimensionless oscillation power $p$ (non-linear power) as function of the current density, and (c) parametric plot of $f$ as function of $p$ for tree different bias fields: 450, 500, and 550 mT. (d) Comparison between micromagnetic and analytical (computed using Eq. 2) linewidth as function of the current density for $H$=500 mT and $T$=300 K, the only fitting parameter is the dimensionless nonlinear frequency shift.

## V. INJECTION LOCKING

In agreement with general synchronization theories, our computation results show the coupling of the auto-oscillation state of a STO to a weak microwave field gives rise to several phenomena (i) frequency locking, (ii) frequency pulling and (iii) frequency modulation.[24,47] P-modes are excited in the injection locking region where basically the self-oscillation mode is locked at the field frequency. In the frequency pulling region, the dynamics is characterized by a quasi-periodic dynamical state (Q-modes) where in the power spectrum is observed the weak mode due to the ac field and the self-oscillation mode pulled from its original oscillation frequency.[24] Figure 5(a) summarizes a stability diagram for phase locking as function of the amplitude of external field $h$ for a current density of $1.6 \times 10^8$ A/cm$^2$ and $H$=500 mT (the central frequency is 10.8 GHz). As can be observed, the regions are symmetric with respect to the central frequency in agreement with recent experiment[22] and analytical theories.[19,23]

For $h \geq 2$ mT (we simulated values up to $h$=3 mT) we found numerically the predicted [23] and measured[48] narrow hysteretic Q/P region (not shown). In the system we studied, this region disappears (or it is smaller than 2 MHz) in the presence of thermal fluctuations. For $h$=2 and 3 mT, we found an hysteretic region of 10 and 20 MHz with a symmetric configuration[23] as expected to have in the range of current density with a linear relationship between oscillation frequency and power.

We also compare our data with the non-linear analytical theory developed in Ref.[19] based on the universal oscillator model. The force locking $f_e$ for an ac field is a complex number given by:

$$f_e = \gamma \left( h_x + i h_y \right) / \sqrt{2} \qquad (3)$$

The bandwidth locking $\Delta$ is related to the force locking by the equation:

$$\Delta = 2\sqrt{1+\nu^2} \frac{F_e}{\sqrt{p(J)}} \qquad (4)$$

being $F_e = |f_e|$, while $p$ and $\nu$ are the oscillation power and the dimensionless non-linear frequency shift computed from the free running data (see Fig. 4). Figure 5(b) shows the comparison among micromagnetic computations of $\dfrac{d\Delta}{2dh}$ (slope of the line which separates the "injection locking" and the "frequency pulling" regions) at $T$=0 K (solid circle) and $T$=300 K (stars) and the non-linear analytical theory (solid line) for $J<1.6 \times 10^8$ A/cm$^2$ (we estimated a $\dfrac{d\Delta}{2dh} = \gamma \dfrac{\sqrt{1+\nu^2}}{\sqrt{p}} = 80 MHz/mT$ considering $\left(1+\nu^2\right) = 2.5$ ). Our results show that this quantitative agreement is achieved up to current densities $J<1.8 \times 10^8$ A/cm$^2$, range of current densities with a linear relationship between $f$ and $p$ (see Figs. 4(a)-(c)).

On the other hand, it can be observed a qualitative prediction of the $\Delta$ by means of Eq. 4 in the whole range of current density where the magnetization dynamics is characterized by a single-mode as displayed in Fig. 5(c), in fact if the oscillation power $p$ increases, $\Delta$ decreases and vice versa and at the minimum in the $\Delta$ corresponds a maximum in the $p$.

Finally, our numerical experiment also shows the possibility to have frequency locking in the 2$^{nd}$ and 3$^{rd}$ harmonic as summarized in the simplified Arnold tongue diagram of Fig. 5(d) (the locking bandwidth is symmetric with respect to the central oscillation frequency). For the system we have studied, the locking bandwidth Δ) in 2$^{nd}$ and 3$^{rd}$ harmonic maintains a linear functional dependence on the force locking even if those are reduced compared to the one of the 1$^{st}$ harmonic. Our findings are in qualitative agreement with a very recent experimental data where, together to the locking at integer harmonics, it has also been measured fractional and integer locking in the regime of moderate and large amplitude of external ac field.[49] Our results point out the possibility to use the STOs also for a different application namely injection-locked frequency dividers.[50]

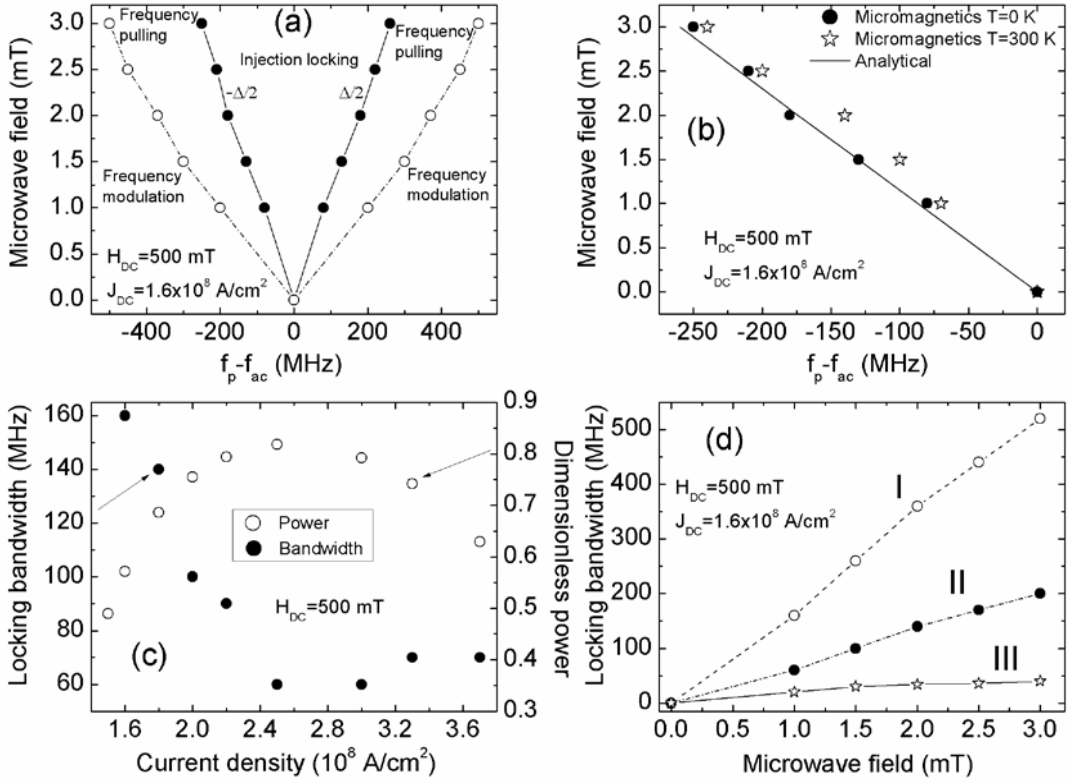

FIG. 5. (a) Stability diagram as function of the amplitude of external field *h* for a current density of 1.6x10$^8$ A/cm$^2$ and *H*=500 mT at *T*=0 K. (b) Micromagnetic stability diagram for the injection locking (symbol) region for *T*=0 K (solid circle) and *T*=300 K (stars) compared to the computation of $\frac{d\Delta}{2dh}$ by means of Eq. 4 (solid line), $f_p$=10.8 GHz. (c) Locking bandwidth and dimensionless oscillation power as function of the current density for *H*=500 mT and *T*=0K. (d) Locking bandwidth for integer harmonics 1$^{th}$ (open circle), 2$^{nd}$ (solid circle), and 3$^{rd}$ (stars) as function of *h* (*J*=1.6x10$^8$ A/cm$^2$)

## VI. SUMMARY AND CONCLUSIONS

We have described, by means of a numerical experiment, the effects of a "weak" microwave field in the fast magnetization reversal and self-oscillation state of spin valves with perpendicular materials.

We find out resonant switching as the field frequency approaches the frequency of the main pre-switching excited mode. The advantage from the practical point of view, with respect to the resonant switching driven by an ac current, is the weakly dependence of the switching time on the relative phase between the ac field and the instant at which the dc current pulse is applied. This gives the possibility to have a more simple scheme for resonant switching magnetoresistive random access memories.

Regarding the injection locking data, our numerical study underlines the possibility to predict the locking bandwidth, for a current region also quantitatively, from the free running data using a non-linear analytical formulation based on the universal model of non-linear oscillators. In other words, we identify the analytical model from the micromagnetic computations of the oscillation power and dimensionless non-linear frequency shift to have an estimation of the locking bandwidth expected.

We also observe the injection locking phenomenon for integer harmonics of the self-oscillation frequency, $2^{nd}$ and $3^{rd}$ harmonic, where the locking bandwidth exhibits a linear dependence on the force locking. This harmonic-injection-locked oscillator can be used as an injection-locked frequency divider.

### ACKNOWLEDGMENTS

This work was supported by Spanish Project under Contracts No. MAT2008-04706/NAN and No. SA025A08. The authors would like to thank Massimiliano D'Aquino for helpful discussions and Sergio Greco for his support with this research.